\documentclass[10pt,a4paper,dvipsnames,usenames]{article}
\usepackage{amssymb}
\usepackage{amsmath}
\usepackage{amsthm}
\usepackage{latexsym}
\usepackage[dvips]{epsfig}
\usepackage{mathrsfs}
\usepackage{bm}
\usepackage{stmaryrd}
\usepackage{authblk}

\usepackage[dvipsnames]{xcolor}
\usepackage{tikz}
\usepackage{enumitem}
\usepackage{commath}
\usepackage{tipa}

\theoremstyle{plain}
\newtheorem{proposition}{Proposition}
\newtheorem{lemma}{Lemma}
\newtheorem{theorem}{Theorem}
\newtheorem{assumption}{Assumption}

\newtheorem*{main}{Theorem}

\newtheorem{remark}{Remark}

\allowdisplaybreaks

\def\bma{{\bm a}}
\def\bmb{{\bm b}}
\def\bmc{{\bm c}}
\def\bmd{{\bm d}}
\def\bme{{\bm e}}

\def\bmg{{\bm g}}

\def\bmo{{\bm o}}

\def\bmzero{{\bm 0}}
\def\bmone{{\bm 1}}

\def\bmA{{\bm A}}
\def\bmB{{\bm B}}
\def\bmC{{\bm C}}

\def\bmepsilon{{\bm \epsilon}}
\def\bmeta{{\bm \eta}}

\def\bmomega{{\bm \omega}}

\def\bmsigma{{\bm \sigma}}

\def\bmtau{{\bm \tau}}

\def\sdiv{\text{\texthtd}}
\def\scurl{\text{\texthtc}}
\def\stwist{\text{\texthtt}}

\def\bmpartial{{\bm \partial}}

\usepackage{mathtools}
\usepackage{xparse}
\makeatletter
\newcommand{\raisemath}[1]{\mathpalette{\raisem@th{#1}}}
\newcommand{\raisem@th}[3]{\raisebox{#1}{$#2#3$}}
\makeatother
\NewDocumentCommand{\newrbar}{O{0pt} O{0pt}}{
  \ensuremath{\mathrlap{\raisemath{#2}{\hspace*{#1}{\mathchar'26\mkern-9mu}}}r}}

\newcounter{mnotecount}

\newcommand{\mnotex}[1]
{\protect{\stepcounter{mnotecount}}$^{\mbox{\footnotesize $\bullet$\themnotecount}}$ 
\marginpar{
\raggedright\tiny\em
$\!\!\!\!\!\!\,\bullet$\themnotecount: #1} }

\newcounter{mnote}

\usepackage{color}
\usepackage{xcolor}
\usepackage[normalem]{ulem}
\usepackage{soul}
\usepackage{hyperref}

\begin{document}

\title{\textbf{Asymptotic charges for spin-1 and spin-2 fields at the
    critical sets of null infinity}}
 
\author[1]{ Mariem Magdy Ali Mohamed \footnote{E-mail
    address:{\tt m.m.a.mohamed@qmul.ac.uk}}}
\author[1]{Juan A. Valiente Kroon \footnote{E-mail address:{\tt j.a.valiente-kroon@qmul.ac.uk}}}

\affil[1]{School of Mathematical Sciences, Queen Mary, University of London,
Mile End Road, London E1 4NS, United Kingdom.}

\maketitle

\begin{abstract}
The asymptotic charges of spin-1 and spin-2 fields are studied near spatial infinity. We evaluate the charges at the critical sets where spatial infinity meets null infinity with the aim of finding the relation between the charges at future and past null infinity. To this end, we make use of Friedrich’s framework of the cylinder at spatial infinity to obtain asymptotic expansions of the Maxwell and spin-2 fields near spatial infinity, which are fully determined in terms of initial data on a Cauchy hypersurface. Expanding the initial data in terms of spin-weighted spherical harmonics, it is shown that only a subset of the initial data, that satisfies certain regularity conditions, gives rise to well-defined charges at the point where future (past) infinity meets spatial infinity. Given such initial data, the charges are shown to be fully expressed in terms of the freely specifiable part of the data. Moreover, it is shown that there exists a natural correspondence between the charges defined at future and past null infinity.
\end{abstract}

\section{Introduction}
\label{Section:Introduction}
The asymptotic structure of gravitational fields in General Relativity can be studied at past/future null infinities or at spatial infinity. Within this context, the subject of asymptotic symmetries in General Relativity has gained increased interest in recent years, due to its relation to soft theorems \cite{SteWein65,Strom2014,TVPA15}, gravitational memory effect \cite{MarcFav10,DemChrit91,LucThi92} and black hole physics \cite{HawPerStro16,HawPerStro17}. While the subject of asymptotic symmetries is abstract and theoretical in nature, the association with the gravitational memory effect provides a possible link to observations. One of the key concepts for studying the asymptotic symmetries in General Relativity is that of conformal compactification, in which points at infinity in physical spacetimes are mapped to points at finite distances. In the standard conformal compactification of Minkowski spacetime, points at infinite null separation are mapped to a smooth null hypersurface, \emph{null infinity}, part of the conformal boundary. Null infinity consists of two disjoint sets: $\mathscr{I}^+$ and $\mathscr{I}^-$, the future and past null infinity, respectively. Moreover, the future and past null infinities are joined by spatial infinity point $i^0$. Penrose's notion of \emph{asymptotic simplicity} is concerned with identifying spacetimes with an asymptotic structure similar to that of Minkowski near null infinity. In particular, a spacetime is said to be asymptotically simple if a null hypersurface $\mathscr{I}$ (null infinity) with two disjoint sets: $\mathscr{I}^+$ and $\mathscr{I}^-$,
can be attached to the spacetime, and all null geodesics acquire two distinct endpoints on the sets $\mathscr{I}^+$ and $\mathscr{I}^-$---see e.g. \cite{PenRin86,CFEBook} for a precise definition. 

\medskip In the null portion of the conformal boundary, it was expected that the asymptotic symmetry group for asymptotically simple spacetimes was to be given by the Poincar\'e group ---the symmetry
group of Minkowski spacetime. However, it is now established that the asymptotic symmetry group for asymptotically simple spacetimes is given by the infinite-dimensional symmetry group, the Bondi-Metzner-Sachs (BMS) group \cite{BMS62, Sachs62}. The infinite number of supertranslations distinguishes the BMS group from the Poincar\'e group. Supertranslations are angle-dependent translations along future
or past null infinity. The BMS symmetry group gives rise to the BMS charges, these charges are generally not integrable within the context of full gravity theory \cite{WaldZou2000,BarTro11}. However, the prescription in \cite{WaldZou2000} allows one to obtain a definition for the conserved charges associated with asymptotic symmetries at null infinity. 

\medskip
The study of the asymptotic symmetries in the spatial regime is more challenging. This is partly due to the singular nature of the conformal structure near spatial infinity for spacetimes with non-vanishing Arnowitt-Deser-Misner (ADM) mass. Different representations for spatial infinity have been proposed in the literature to deal with this singular behaviour e.g. \cite{AshHan78, AshRom92, Fri98a}. The asymptotic symmetry group for General Relativity at spatial infinity had been studied in \cite{Geroch72,AshHan78,Ash80,AshRom92,MarcCedric18,MarcCedric182}. Similar to the null case, it turns out that one also obtains an infinite-dimensional asymptotic symmetry group, the Spi-group \cite{AshHan78, AshRom92}, and corresponding conserved charges associated with these symmetries. The structure of the Spi-group turns out to be similar to the BMS group. More specifically, the Spi-group is the semi-direct product of the Lorentz group with the Spi-supertranslations \cite{AshHan78}.

\medskip
An interesting problem that arises when looking at the asymptotic symmetries is the so-called matching problem ---i.e. the problem of the relation between the asymptotic symmetries at past and future null
infinity in the limit of spatial infinity. In \cite{Strom2014}, Strominger proposed the conjecture stating that the BMS symmetries at past and future null infinity can be related by antipodal reflection
near spatial infinity. The main technique utilised in proving the matching problem is to make use of the field equations and charges
defined at spatial infinity to relate the charges at past and future null infinities. On Minkowski spacetime, the matching problem had
been investigated for the spin-1 field (Maxwell field) in \cite{CamEyh17,MarcCedric183} and for the spin-2 field (linearised gravity) in \cite{Troe18}. While on Minkowski spacetime, the transformation
between the adapted asymptotic coordinates at spatial infinity \cite{BeiSch82} and null infinity \cite{Bondi60, BMS62} can be derived explicitly, the relation between the asymptotic coordinates for
general spacetimes is not explicitly known. This, in addition to the singular nature of spatial infinity, makes the matching problem more complicated in more general spacetimes. One way to address this is to make use of a covariant formulation in which the singular nature of spatial infinity is resolved. In \cite{Kartik18,Kartik19}, Ashtekar and Hansen's formulation of asymptotic flatness was used to show that the matching
of supertranslation charges holds for spin-1 and gravitational fields in general asymptotically flat spacetimes. In this instance, the notion of asymptotic flatness is the one introduced in \cite{AshHan78}. More recently, the matching problem has been investigated for Lorentz charges in General Relativity for spacetimes that satisfy Ashtekar and Hansen's notion of asymptotic flatness
\cite{KarIbr2021}. That being said, in this paper, we revisit the asymptotic symmetries and matching problem for spin-1 and spin-2 fields on Minkowski spacetime. In contrast to \cite{CamEyh17, KartikIbrahim20, MarcCedric183, MarcCedric19}, our aim is to express the asymptotic charges at the critical sets in terms of the freely specifiable initial data on a Cauchy hypersurface. To do so, we use a different formulation for spatial infinity ---namely, Friedrich's cylinder at spatial infinity. \emph{The advantage of using Friedrich's formulation is that it allows us to directly link the charges defined
at past and future null infinity to initial data prescribed on a Cauchy hypersurface.} The approach followed in this paper will serve as a stepping stone to investigate the asymptotic charge and the
matching problem for the full theory of gravity in Friedrich's formalism in later work.

\medskip
Friedrich's conformal representation of spatial infinity was first introduced in \cite{Fri98a} with the aim of formulating a regular initial value problem for the conformal Einstein field equations. In this representation, the spatial infinity point is blown up to $(-1,1) \times \mathbb{S}^2$, which is known as the cylinder at spatial infinity $\mathcal{I}$. The end 'points' of the cylinder, known as the
\emph{critical sets} $\mathcal{I}^{\pm} = \{ \pm 1 \} \times \mathbb{S}^2$, are the set of points at which the future and past null infinity touches the cylinder $\mathcal{I}$. This representation of spatial infinity proves to be useful in relating quantities at $\mathcal{I}^{\pm}$ to initial data given on a Cauchy hypersurface. This framework had been used to obtain expressions for the Newman-Penrose (NP) constants in terms of initial data in \cite{FriKan00}. More recently, it was shown that, generically, there is no natural correspondence between the NP constants for spin-1 and spin-2 fields on $\mathscr{I}^+$ and $\mathscr{I}^-$ for generic initial data \cite{GasVal20}, except for the particular case of time-symmetric initial data.

\medskip
In the literature, the asymptotic charges are usually given in terms of the NP gauge. For the purpose of this article, we make sure of frames that are more suited for the framework of the cylinder at spatial infinity, the so-called F-gauge. These frames are adapted to a congruence of conformal geodesics and Cauchy hypersurfaces ---see \cite{FriKan00}. For Minkowski spacetime, the relation between the NP gauge and F-gauge is given explicitly in \cite{GasVal20}. One can then rewrite the expressions for the asymptotic charges in terms of field components in the F-gauge. Using the expansion given in Assumption \ref{Maxwell-field-expansion} in the Maxwell case and given a generic class of initial data, we show that the charges associated with supertranslations at $\mathcal{I}^+$ and at $\mathcal{I}^-$ can be written in terms of freely specifiable initial data on the initial hypersurface $\mathcal{S}_\star \equiv \{ \tau=0 \}$. These charges are not in general well-defined, and one requires extra conditions ---see Lemma \ref{Maxwell-conditions}, for well-defined charges at the critical sets $\mathcal{I}^{\pm}$. Finally, we show that there is a natural correspondence between the charges at $\mathcal{I}^+$ and $\mathcal{I}^-$ in the sense that they can be traced back to the same piece of freely specifiable data on the initial Cauchy hypersurface. Similarly, in the spin-2 case, the expansion given in Assumption \ref{spin-2-field-expansion} and the generic initial data are used to show that the charges associated with supertranslations at the critical sets can be written in terms of initial data on the initial hypersurface. The conditions given in Lemma \ref{Spin2-conditions}
ensures that the charges are well-defined at $\mathcal{I}^{\pm}$.

\medskip
As will be seen in the main text of the article, in order to obtain non-vanishing BMS charges for both the Maxwell and spin-2 fields, one needs to consider initial data sets which are \emph{boosted} ---that is, the leading terms of the various components of the initial data contain spin-weighted harmonics ${}_n Y_{l,m}$ with $l\geq 1$. For non-boosted initial data sets, only the $l=0$ harmonic gives a non-trivial BMS charge.  The precise form of the notion of boosted data for the Maxwell and spin-2 fields used in this article are given in equations \eqref{Maxwell-initial-data} and \eqref{spin2-initial-data}, respectively. 


\subsection*{Main results of this article}

The main results of the analysis in this article can be summarised in the following:

\begin{main} 
For generic boosted initial data for the Maxwell and spin-2 fields, the asymptotic charges at $\mathcal{I}^{\pm}$ are well-defined if and only if the freely specifiable data on the initial hypersurface $\mathcal{S}_\star$ satisfy certain regularity conditions ---given by Lemma \ref{Maxwell-conditions}; Lemma \ref{Spin2-conditions}, respectively. Moreover, for a given harmonic
$Y_{l,m}$ the charges at $\mathcal{I}^+$ and $\mathcal{I}^-$ are expressible in terms of the same piece of freely specifiable data ---thus, in this sense, they can be identified.
\end{main}

The detailed formulation of the results in the above theorem is given in Theorems \ref{Theorem:ChargesMaxwell} and \ref{Theorem:ChargesSpin2}.


\subsection*{Outline of the article}
This article is structured as follows. In Section \ref{Section:Cylinder}, we introduce the construction of the cylinder at spatial infinity and the F-gauge on Minkowski spacetime. The Newman-Penrose
gauge conditions and the relation between the NP-gauge and the F-gauge for Minkowski spacetime is summarised in Section \ref{Section:NP}. In Section \ref{Section:Maxwell}, the Maxwell equations are expanded in the F-gauge, the initial data for the Maxwell field are discussed, Finally, the asymptotic charges associated with supertranslations are written in terms of initial data given on a spacelike hypersurface. In
Section \ref{Section:Spin2}, we follow a similar strategy to expand the spin-2 field equation using the F-gauge and write the asymptotic charges associated with supertranslations in terms of initial data.

\subsection*{Notations and conventions}
In what follows, the Latin letters $a,\,b,\,c,\ldots$ will denote spacetime abstract tensorial indices, while $A,\,B,\,C,\ldots$ denote abstract spinorial
indices. The Greek letter $\mu,\,\nu,\dots$ are used as spacetime coordinate indices while $\alpha,\, \beta,\ldots$ are used as spatial coordinate indices. Boldface Latin letters $\bma, \bmb, \bmc, \ldots$ are used as tensorial frame indices and boldface capital Latin letters $\bmA, \bmB, \bmC, \ldots$ are used as spinorial frame indices. Using these notations, given a basis $\bme_{a} \equiv \{ \bme_{\bma} \}$, the components of a generic tensor $T_{ab}$ with respect to the basis $\{ \bme_{\bma} \}$ can be written as $T_{\bma \bmb} = T_{ab} \bme_{\bma}{}^a{} \bme_{\bmb}{}^b{}$. Similarly, given a spin basis $\{ o, \iota \}$ that satisfy $\llbracket o, \iota \rrbracket =1$, where $\llbracket .,.\rrbracket$ is the antisymmetric product, the components of a generic spinor $\zeta_{A}$ can be written as $\zeta_{\bmA} = \zeta_{A} \bmepsilon_{\bmA}{}^A{}$ where 
\begin{eqnarray*}
&& \bmo^A = \bmepsilon_{\bmzero}{}^A{} \hspace{2mm}, \hspace{2mm}  \iota^A = \bmepsilon_{\bmone}{}^A{} \\
&& \bmo_A = \bmepsilon^{\bmone}{}_A{} \hspace{2mm} , \hspace{2mm} \iota_A = - \bmepsilon^{\bmzero}{}_A{}.
\end{eqnarray*}
The antisymmetric product $\llbracket .,. \rrbracket$ of two generic spinors can be expressed as 
$$
\llbracket \zeta, \lambda \rrbracket = \zeta_B \lambda^B = \bmepsilon_{AB} \zeta^A \lambda^B,
$$
where $\bmepsilon_{AB}$ is known as the $\bmepsilon$-spinor, and it is an antisymmetric spinor that can be regarded as an index raising/lowering object for spinors. In the rest of the article, we express spacetime frames $\{ \bme_{\bma} \}$ in spinorial notation. The spinorial counterpart of $\{ \bme_{\bma} \}$ is given by
$$
\bme_{\bmA \bmA'} = \sigma^{\bma}{}_{\bmA \bmA'} \bme_{\bma}
$$
where $ \sigma^{\bma}{}_{\bmA \bmA'}$ are the Infeld-van der Waerden symbols. Finally, the signature convention for spacetime metrics used in this article is $(+,-,-,-)$. Throughout, we mostly follow the notation and conventions of Penrose \& Rindler \cite{PenRin86} ---see also \cite{CFEBook}.


\section{The cylinder at spatial infinity and the F-gauge}
\label{Section:Cylinder}
In this section, we introduce the conformal representation of Minkowski spacetime, which gives rise to Friedrich's cylinder at spatial infinity. In this conformal representation, spatial infinity $i^0$ is blown up to $\mathbb{R} \times \mathbb{S}^2$. In the following, we will only introduce the cylinder at spatial infinity for Minkowski spacetime. A more general discussion of this representation is presented in the original paper \cite{Fri98a}.


\subsection{The cylinder at spatial infinity}
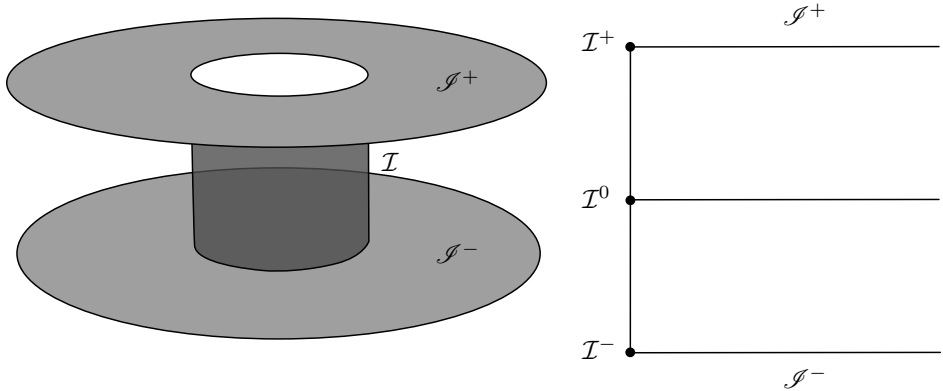
\begin{figure*}
\includegraphics[width=\textwidth]{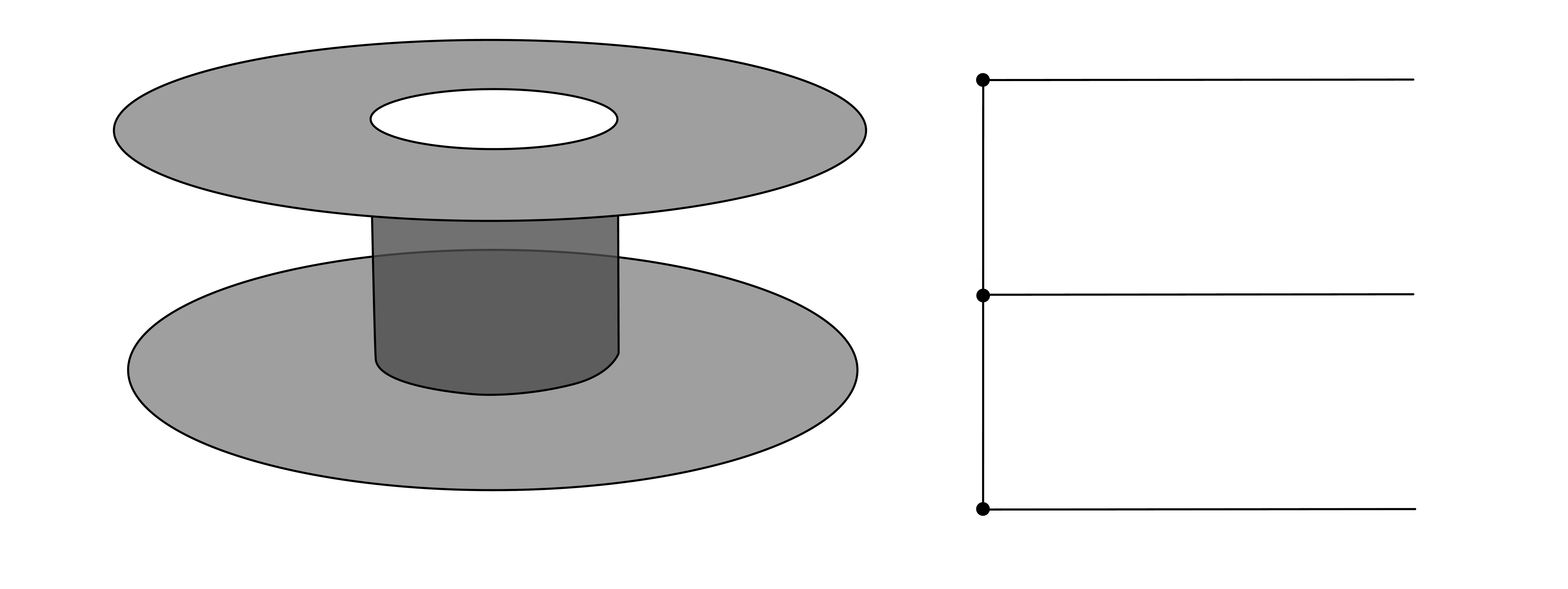}
\put(-175,135){$\mathcal{I}^+$}
\put(-175,76){$\mathcal{I}^0$}
\put(-175,19){$\mathcal{I}^-$}
\put(-100,145){$\mathscr{I}^+$}
\put(-100,8){$\mathscr{I}^-$}
\put(-250,90){$\mathcal{I}$}
\put(-230,120){$\mathscr{I}^+$}
\put(-230,55){$\mathscr{I}^-$}
\caption{Schematic representation of the sets of the cylinder at spatial infinity. Note that the two-dimensional representation on the right is not a conformal diagram.}
\label{Fig:Cylinder}
\end{figure*}

The Minkowski metric written in Cartesian coordinates $(\tilde{t},\tilde{x}^\alpha)$ is given by
\[
\tilde{\bmeta} = \eta_{\mu\nu} \bmd{\tilde{x}^{\mu}} \otimes \bmd{\tilde{x}^{\nu}},
\]
where $\eta_{\mu\nu} = \text{diag}\left( 1, -1, -1, -1 \right)$. Defining $\tilde{r}^2 \equiv \delta_{\alpha\beta} \tilde{x}^\alpha \tilde{x}^\beta$, the metric can be written in polar coordinates as
\[
\tilde{\bmeta} = \bmd{\tilde{t}} \otimes \bmd{\tilde{t}} - \bmd{\tilde{r}} \otimes \bmd{\tilde{r}} - \tilde{r}^2 \bmsigma,
\]
where $\bmsigma$ is the standard metric on $\mathbb{S}^2$. Now introduce the inversion coordinates $(t,x^\alpha)$ where
\[
x^\mu= -\frac{\tilde{x}^\mu}{\tilde{X}^2}, \qquad \tilde{X}^2\equiv \tilde{\eta}_{\mu\nu} \tilde{x}^\mu \tilde{x}^\nu.
\]
Then one can introduce the unphysical metric $\hat{\bmeta} \equiv X^4
\tilde{\bmeta}$ where $X^2 \equiv \eta_{\mu\nu} x^\mu x^\nu$. Hence, we get
\[
\hat{\bmeta} = \bmd{t} \otimes \bmd{t} - \bmd{\rho} \otimes \bmd{\rho} - \rho^2 \bmsigma,
\]
where $\rho^2 \equiv \delta_{\alpha\beta} x^\alpha x^\beta$ . In this conformal representation, the spatial infinity point $i^0$ is at the
origin. Introduce $\tau \equiv t/ \rho$ and define
\begin{equation}
    \bmeta \equiv  \frac{1}{\rho^2} \hat{\bmeta} = \frac{X^2}{\rho^2} \tilde{\bmeta}.
    \label{Minkowski-metric-Friedrich-coordinates}
\end{equation}
Then, one has that
\begin{equation}
    \bmeta = \bmd{\tau} \otimes \bmd{\tau} - \frac{(1-\tau^2)}{\rho^2} \bmd{\rho} \otimes \bmd{\rho} + \frac{\tau}{\rho} \left( \bmd{\tau} \otimes \bmd{\rho} + \bmd{\rho} \otimes \bmd{\tau} \right) - \bmsigma.
\end{equation}
This metric gives the desired representation of spatial infinity. In particular, spatial infinity corresponds to a set with the topology of $\mathbb{R} \times \mathbb{S}^2$. Consider the conformal extension
$(\mathcal{M}, \bmeta)$ where $\bmeta$ is given by equation
\eqref{Minkowski-metric-Friedrich-coordinates} and
\[
\mathcal{M} \equiv \{ p \in \mathbb{R}^4 \ | \ -1 \leq \tau(p) \leq 1,
\ \rho(p) \geq  0 \}.
\]
Then, the future and past null infinities are located at 
\[
\mathscr{I}^\pm \equiv \big\{ p\in \mathcal{M} \ | \  \tau(p) =\pm 1 \big\}.
\]
Moreover, the following sets can be introduced 
\begin{eqnarray*}
&& \mathcal{I} \equiv \big\{ p \in \mathcal{M} \ | \ |\tau(p)|<1, \ \rho(p)=0\big\}, \nonumber \\
&& \mathcal{I}^{0} \equiv \big\{ p \in \mathcal{M} \ |\  \tau(p)=0, \ \rho(p)=0\big\},
\end{eqnarray*}
and 
\begin{eqnarray*}
&& \mathcal{I}^{+} \equiv \big\{ p\in \mathcal{M} \ | \ \tau(p)=1, \ \rho(p)=0 \big\},  \nonumber\\
&& \mathcal{I}^{-} \equiv \big\{p \in \mathcal{M} \ | \ \tau(p)=-1, \  \rho(p)=0\big\},
\end{eqnarray*}
where $\mathcal{I}$ is the cylinder at spatial infinity, $\mathcal{I}^{\pm}$ are the critical sets at which spatial infinity touches $\mathscr{I}^+$ and $\mathscr{I}^-$. Finally, $\mathcal{I}^0$
is the intersection of $\mathcal{I}$ with the initial hypersurface $\mathcal{S}_\star \equiv \{ \tau = 0 \}$. A schematic representation of these sets is shown in Figure  \ref{Fig:Cylinder}.

\begin{remark}
{\em In this conformal representation, null infinity coincides with the hypersurfaces given by the condition $\tau=\pm1$. However, this can be generalised so that future and past null infinity do not
correspond to $\tau = \pm 1$ by considering the transformation $\tau \mapsto \tau/ \kappa$ where $\kappa = \varkappa \rho$ and $\varkappa= O(\rho^0)$. For simplicity, we choose $\varkappa = 1$ for the rest of
the paper. This representation with $\varkappa =1$ is known as the F-gauge \emph{horizontal representation}. }
\end{remark}


\subsection{F-gauge}
As mentioned in the introduction of this article, we will make use of the F-gauge frames, which are more suited for the framework of the cylinder at spatial infinity, to write the asymptotic charges for spin-1 and spin-2 fields in terms of initial data. To construct these frames, let $\mathcal{Q}_{\tau,\rho}$ denote the 2-dimensional surfaces in $\mathcal{M}$ of
constant $\rho$ and $\tau$. Then the metric on $\mathcal{Q}_{\tau,\rho}$ is the standard
metric on $\mathbb{S}^2$. One can introduce a complex null frame $\{
\bmpartial_+, \bmpartial_- \}$ on $\mathcal{Q}_{\tau,\rho}$ and impose
\[
[ \bmpartial_{\tau}, \bmpartial_{\pm} ] =0, \qquad [ \bmpartial_{\rho}, \bmpartial_{\pm} ]=0.
\]
These conditions allows us to propagate $\{ \bmpartial_+, \bmpartial_-
\}$ off $\mathcal{Q}_{\tau\rho}$. Now, introduce the frames
\cite{GasVal20}
\begin{eqnarray*}
&& \bme_{\bmzero \bmzero'} = \frac{\sqrt{2}}{2} \left( (1-\tau) \bmpartial_{\tau} + \rho \bmpartial_{\rho} \right), \nonumber\\
&& \bme_{\bmone \bmone'} = \frac{\sqrt{2}}{2} \left( (1+\tau) \bmpartial_{\tau} - \rho \bmpartial_{\rho} \right), \nonumber\\
&& \bme_{\bmzero \bmone'} = \frac{\sqrt{2}}{2} \bmpartial_+, \nonumber\\
&& \bme_{\bmone \bmzero'} = \frac{\sqrt{2}}{2} \bmpartial_-.
\end{eqnarray*}
and the dual frames
\begin{eqnarray*}
&& \bmomega^{\bmzero \bmzero'} = \frac{\sqrt{2}}{2} \left( \bmd{\tau} - \frac{1}{\rho} (1-\tau) \bmd{\rho} \right), \nonumber\\
&& \bmomega^{\bmone \bmone'} = \frac{\sqrt{2}}{2} \left( \bmd{\tau} + \frac{1}{\rho} (1+\tau) \bmd{\rho} \right), \nonumber\\
&& \bmomega^{\bmzero \bmone'} = \sqrt{2} \bmomega^+,  \nonumber\\
&& \bmomega^{\bmone \bmzero'} = \sqrt{2} \bmomega^-,
\end{eqnarray*}
where $\bmomega^\pm$ are dual to $\bmpartial_\pm$ ---that is, one has
\[
\langle \bmomega^+,\bmpartial_+\rangle =1, \qquad \langle \bmomega^-,\bmpartial_-\rangle =1.
\]
In terms of the above fields, the metric $\bmeta$ can be written as
$$
\bmeta = \bmepsilon_{\bmA \bmB} \bmepsilon_{\bmA' \bmB'} \bmomega^{\bmA \bmA'} \otimes \bmomega^{\bmB \bmB'}.
$$


\section{The Newman-Penrose gauge}\label{Section:NP}
In this section, we briefly summarise NP-gauge conditions satisfied by NP frames as well as the relation between the F-gauge frames and the null frames in Minkowski spacetime derived in \cite{GasVal20}. In the NP gauge, the frame fields are adapted to null infinity ---the conditions listed here apply for NP frames at $\mathscr{I}^+$. Analogous conditions can be formulated, \emph{mutatis mutandi}, for $\mathscr{I}^-$. Let $\{ \bme'_{\bmA \bmA'} \}$ denote a frame satisfying $\bmeta(\bme'_{\bmA \bmA'}, \bme'_{\bmB \bmB'})=\bmepsilon_{\bmA \bmB} \bmepsilon_{\bmA' \bmB'}$ in a neighbourhood $\mathcal{U}$ of $\mathscr{I}^+$. One requires a null frame in the NP-gauge to satisfy the following conditions:
\begin{itemize}
    \item[(i)] $\bme'_{\bmone \bmone'}$ is tangent to $\mathscr{I}^+$ and satisfies 
    $$
    \nabla_{\bmone \bmone'} \bme'_{\bmone \bmone} =0.
    $$

    \item[(ii)] There exits a smooth function $u$ on $\mathcal{U}$ that satisfies $\bme'_{\bmone \bmone'}(u) = 1$ at $\mathscr{I}$.

    \item[(iii)] The frame $\bme'_{\bmzero \bmzero'}$ is defined as 
    $$
    \bme'_{\bmzero \bmzero'} = \bmeta^{\sharp}(\bmd{u},\cdot).
    $$
    \item[(iv)] Define 
    $$
    \mathcal{N}_{u_\bullet} \equiv \{ p \in \mathcal{U} \ | \ u(p) = u_\bullet \},
    $$
    where $u_\bullet$ is constant. Then one requires the frame $\bme'_{\bmA \bmA'}$, tangent to $\mathcal{N}_{u_\bullet} \cup \mathscr{I}^+$, to satisfy
    $$
    \nabla_{\bmzero \bmzero'} \bme'_{\bmA \bmA'} = 0 \hspace{2mm} \text{on} \hspace{2mm} \mathcal{N}_{u_\bullet}.
    $$
\end{itemize}

In \cite{GasVal20}, the relation between the NP-gauge frame $\{ \bme'_{\bmA \bmA'} \}$ and the F-gauge frame$\{ \bme_{\bmA \bmA'} \}$ for Minkowski spacetime was explicitly computed. This computation is summarised in the following:

\begin{proposition}
The NP-gauge frame at $\mathscr{I}^+$ and F-gauge frame on Minkowski spacetime are related via 
\begin{equation}
    \bme'_{\bmA \bmA'} = \Lambda^{\bmB}{}_{\bmA} \bar{\Lambda}^{\bmB'}{}_{\bmA'} \bme_{\bmB \bmB'},
    \label{NP-FGauge-Relation}
\end{equation}
and 
\begin{eqnarray}
&& \Lambda^{\bmone}{}_{\bmzero} = \frac{2 e^{i \omega}}{\sqrt{\rho}(1+\tau)}, \hspace{5mm} \Lambda^{\bmzero}{}_{\bmone} = \frac{e^{-i \omega} \sqrt{\rho} (1+\tau)}{2},\nonumber\\
&& \Lambda^{\bmone}{}_{\bmone} = \Lambda^{\bmzero}{}_{\bmzero} =0,
\end{eqnarray}
where $\omega$ is an arbitrary real number that encodes the spin rotation of the frames on $\mathbb{S}^2$. For the NP-gauge frame at $\mathscr{I}^-$, the roles of the vectors $\bme'_{\bmzero \bmzero'}$ and $\bme'_{\bmone \bmone'}$ are interchanged, and the NP-gauge frame is related to the F-gauge by equation \eqref{NP-FGauge-Relation} with $\Lambda^{\bmA}{}_{\bmB}$ given by
\begin{eqnarray}
&&  \Lambda^{\bmone}{}_{\bmzero} = \frac{e^{-i \omega} \sqrt{\rho} (1-\tau)}{2} , \hspace{5mm} \Lambda^{\bmzero}{}_{\bmone} = \frac{2 e^{i \omega}}{\sqrt{\rho}(1-\tau)},\nonumber \\ 
&& \Lambda^{\bmone}{}_{\bmone} = \Lambda^{\bmzero}{}_{\bmzero} =0.
\end{eqnarray}

\end{proposition}


\section{The Maxwell field and asymptotic charges in the F-gauge}
\label{Section:Maxwell}

In this section, we discuss the Maxwell field on Minkowski spacetime. First, we start by writing the Maxwell equations in the F-gauge and expanding the Maxwell field in terms of spin-weighted
spherical harmonics. We then use the Maxwell equations to solve for the coefficients of the expansion. The analysis here follows the discussion found in \cite{GasVal20}.


\subsection{The spinorial Maxwell equations}
Starting with the Maxwell equations in the 2-spinor formalism 
\begin{equation}
    \nabla_{A'}{}^{A} \phi_{AB} = 0,
    \label{Maxwell-Spinorial}
\end{equation}
applying $- 2 \nabla_C{}^{A'}$, one can readily finds that the spinor $\phi_{AB}$ satisfies the wave equation  
\begin{equation}
   \square \phi_{AB} =0, 
   \label{WaveEqnMaxwell}
\end{equation}
where $\square$ is the D'Alembertian operator $\square \equiv \nabla_{AA'}\nabla^{AA'}$. We define a spin-dyad similar to the frame $\{\bme_{\bmA\bmA'}\}$. Accordingly, define $\bmepsilon_{\bmA}{}^A$ with
$\bmepsilon_\bmzero{}^A = \bmo^A$ and $\bmepsilon_\bmone{}^A = \iota^A$ as the spin-dyad in the F-gauge. Then equation~\eqref{WaveEqnMaxwell} can be split into a system of equations written in terms of the components of $\phi_{AB}$ with respect to $\bmepsilon_\bmA{}^A$, where $\phi_0 \equiv \phi_{AB} \iota^A \iota^B$, $\phi_1 \equiv \phi_{AB} o^A \iota^B$ and $\phi_2 \equiv \phi_{AB} o^A o^B$ so as to obtain the
scalar equations
\begin{subequations}
\begin{equation}
    (1-\tau^2) \ddot{\phi}_0 + 2 \rho \tau \partial_{\rho} \dot{\phi}_0 - \rho^2 \partial^2_{\rho}\phi_0 - \frac{1}{2} (\eth \Bar{\eth} \phi_0 + \Bar{\eth} \eth \phi_0) - 2(1+\tau) \dot{\phi}_0 + \phi_0 =0, \label{MaxwellWaveExpanded1}
\end{equation}
\begin{equation}
    (1-\tau^2) \ddot{\phi}_1 + 2 \rho \tau \partial_{\rho} \dot{\phi}_1 - \rho^2 \partial^2_{\rho}\phi_1 - \frac{1}{2} (\eth \Bar{\eth} \phi_1 + \Bar{\eth} \eth \phi_1) + 2 \tau (\rho -1) \dot{\phi}_1 = 0, \label{MaxwellWaveExpanded2} 
\end{equation}
\begin{equation}
    (1-\tau^2) \ddot{\phi}_2 + 2 \rho \tau \partial_{\rho} \dot{\phi}_2 - \rho^2 \partial^2_{\rho}\phi_2 - \frac{1}{2} (\eth \Bar{\eth} \phi_2 + \Bar{\eth} \eth \phi_2)+ 2 (1-\tau) \dot{\phi}_2 + \phi_2 =0. \label{MaxwellWaveExpanded3}
\end{equation}
\end{subequations}
where $\eth$ and $\bar{\eth}$ are the Newman and Penrose operators \cite{NewPen62, PenRin86, Stewart91}. The relation between the NP operators ($\eth$, $\Bar{\eth}$) and ($\bmpartial_+$, $\bmpartial_-$) is given in Appendix B in \cite{GasVal20}. To analyse the solutions of these equations, the following Ansatz is made:

\begin{assumption}
\label{Maxwell-field-expansion}
The components of the Maxwell field admit an expansion around $\rho=0$ and for $\tau\in[-1,1]$ of the form
 \begin{equation}
     \phi_n = \sum_{l=|n-1|}^{\infty} \sum_{m=-l}^l a_{n;l,m}(\tau) {}_{n-1}Y_{l,m} + o_1(\rho),
 \label{MaxwellFieldExpansion}
 \end{equation}
where $a_{n;l,m}: \mathbb{R} \rightarrow \mathbb{C}$ and $n=0,\,1,\,2$ and  ${}_nY_{l,m}$ are spin-weighted spherical harmonics and the symbol $o_1(\rho)$ indicates that the higher order terms and their first derivatives remain bounded uniformly as $\rho\rightarrow 0$ and $\tau\rightarrow \pm 1$ with the corresponding decay rate ---i.e. $o(\rho)$ for the function and $o(1)$ for the derivative. 
\end{assumption}

\begin{remark}
{\em Making use of the estimates developed in \cite{Fri03b,Val09a} it is possible to show that there exist large classes of solutions to the Maxwell equations having an expansion consistent with \eqref{MaxwellFieldExpansion}. More precisely, it is possible to construct solutions of the form 
  \[
  \phi_n = \sum_{k=|1-k|}^{p-1} \frac{1}{k!}\phi_n^{(k)}\rho^k + R_p[\phi_n],
  \]
where the coefficients $\phi_n^{(k)}$ are explicitly known functions of $\tau$ and the angular variables which are smooth for $\tau\in(-1,1)$ and whose regularity at $\tau=\pm 1$ can be controlled in terms of the initial data. Moreover, the reminder satisfies $R_p[\phi_n]\in C^m$ for $p\geq m+5$ for $\rho$ near 0 and $\tau\in [-1,1]$. In particular, for the expansion in Assumption \ref{Maxwell-field-expansion}, one needs $p\geq 6$. The coefficients $\phi_n^{(k)}$ are determined by solving transport equations on the cylinder $\mathcal{I}$ in an analogous manner to what is done below for the coefficients $a_{n;l,m}(\tau)$ in expansion \eqref{Maxwell-field-expansion}.  }
\end{remark}


Using equation \eqref{MaxwellFieldExpansion} and substituting in equations \eqref{MaxwellWaveExpanded1}-\eqref{MaxwellWaveExpanded3}, one readily finds that the coefficients $a_{n;l,m}$ satisfy the equations
\begin{subequations}
\begin{equation}
    (1-\tau^2) \ddot{a}_{0;l,m} - 2 (1+\tau) \dot{a}_{0;l,m} + l(l+1) a_{0;l,m} = 0,
    \label{WaveEqnExpansion1} \\
\end{equation}
\begin{equation}
    (1-\tau^2) \ddot{a}_{1;l,m} - 2 \tau \dot{a}_{1;l,m} + l(l+1) a_{1;l,m} =0,
    \label{WaveEqnExpansion2} \\
\end{equation}
\begin{equation}
    (1-\tau^2) \ddot{a}_{2;l,m} + 2 (1-\tau) \dot{a}_{2;l,m} + l(l+1) a_{2;l,m} =0. \label{WaveEqnExpansion3}
\end{equation}
\end{subequations}

As will be seen later, the expression for the Maxwell charges is written in terms of the coefficients $a_{1;l,m}$. Thus, we are interested in the general solution of equation \eqref{WaveEqnExpansion2}. This is described by the following proposition:

\begin{proposition}
For $ l \geq 0$ and $-l \leq m \leq l$, the general solution to equation \eqref{WaveEqnExpansion2} is given by
\[
a_{1;l,m} = A_{l,m} P_{l}(\tau) + B_{l,m} Q_{l}(\tau),
\]
where $P_{l}(\tau)$ denotes the Legendre polynomial of order $l$ and $Q_{l}(\tau)$ is the Legendre function of the second kind of order $l$. Moreover, $A_{l,m}$ and $B_{l,m}$ are constants that can be expressed in terms of the initial conditions $a_{1;l,m}(0)$ and $\dot{a}_{1;l,m}(0)$ as
\begin{subequations}
\begin{equation}
    A_{l,m} = -(l+1) Q_{l+1}(0) a_{1;l,m}(0) - Q_{l}(0) \dot{a}_{1;l,m}(0), \label{Maxwell-Alm} 
\end{equation}
\begin{equation}
        B_{l,m} = \frac{\sqrt{\pi} (l+1)}{\Gamma(-\frac{l}{2}) \Gamma(\frac{l+3}{2})} a_{1;l,m}(0) + \frac{\sqrt{\pi}}{\Gamma(\frac{1}{2}-\frac{l}{2}) \Gamma(1+\frac{l}{2})} \dot{a}_{1;l,m}(0). \label{Maxwell-Blm}
\end{equation}

\end{subequations}
\label{Jacobi-solution}
\end{proposition}

\begin{remark}
\label{ReplyReferee}
{\em The analysis in \cite{Fri03b,Val09a} has been restricted to expansions for which the leading term in $\phi_1$ is of the form
\[
a_{1;0,0}(\tau)\, Y_{0,0}.
\]
As will be seen in the following, the higher order harmonics in \eqref{MaxwellFieldExpansion} are required to generate non-trivial generic BMS charges.
}
\end{remark}

\begin{remark}
\label{Remark:Logs}
{\em For non-vanishing values of $B_{l,m}$, one can directly deduce from the recurrence relation of the Legendre function of the second kind that the solution given by Proposition \ref{Jacobi-solution} diverges logarithmically near $\tau = \pm 1$ in the sense that
\[
Q_l(\tau) = \mathcal{C}_l \ln(1 \pm \tau) + O(1), \mbox{for some constant
$\mathcal{C}_l$.}
\]
} 
\end{remark}

From equation \eqref{Maxwell-Blm}, one can deduce that requiring that $B_{l,m} = 0$ gives the following:
\begin{lemma}
\label{Initial-data-Maxwell}
The solution given in Proposition \ref{Jacobi-solution} is regular if and only if 
\begin{itemize}
    \item[(i)] $a_{1;l,m}(0)=0$ for odd $l$, and
    \item[(ii)] $\dot{a}_{1;l,m}(0)=0$ for even $l$.
\end{itemize}
\end{lemma}


\subsection{Initial data for the Maxwell equations}
\label{Section:Initial-data-for-Maxwell}
The coefficients of the expansion given in Assumption \ref{Maxwell-field-expansion} are fully determined by the evolution and constraint equations implied by equation \eqref{Maxwell-Spinorial}. In this section, we express the initial data conditions given in Lemma \ref{Initial-data-Maxwell} in terms of freely specifiable initial data. The space-spinor decomposition of equation \eqref{Maxwell-Spinorial} gives the evolution and constraint equations which can be written, respectively, as
\begin{subequations}
\begin{equation}
    \mathcal{D} \phi_{AB} + 2 \mathcal{D}_A{}^C \phi_{CB} =0, \hspace{6mm} \text{(evolution equations)}
\end{equation}
\begin{equation}
    \mathcal{D}^{AB}\phi_{AB} =0, \hspace{21.5mm} \text{(constraint equation)}
\end{equation}
\end{subequations}
where 
\[
\mathcal{D} \equiv \tau^{AA'} \nabla_{AA'}, \qquad \mathcal{D}_{AB}
\equiv \tau_{(A}{}^{C'} \nabla_{B)C'},
\]
and $\tau_{AA'}$ is the spinorial counterpart of a future directed timelike vector $\bmtau$ satisfying
\[
\tau^A{}_{A'} \tau^{BA'} = \bmepsilon^{AB}, \hspace{2mm} \tau_{AA'} \tau^{AA'} =2.
\]
In particular, $\bmtau$ is orthogonal to the initial hypersurface $\mathcal{S}_\star$ given by $\tau=0$. Expanding in terms of the spin-dyad of the F-gauge, one can write the evolution equations as
\begin{subequations}
\begin{equation}
    (1-\tau) \dot{\phi}_0 + \rho \partial_{\rho}\phi_0 + \Bar{\eth}\phi_1 = \phi_0, \label{Evolution-Maxwell1}\\
\end{equation}
\begin{equation}
    \dot{\phi}_1 + \frac{1}{2} \left( \eth \phi_0 + \Bar{\eth}\phi_2 \right) =0,  \label{Evolution-Maxwell2}\\
\end{equation}
\begin{equation}
    (1+\tau) \dot{\phi}_2 - \rho \partial_{\rho}\phi_2 + \eth \phi_2 = - \phi_2, \label{Evolution-Maxwell3}   
\end{equation}
\end{subequations}
and the constraint equation as
\begin{equation}
        \tau \dot{\phi}_1 - \rho \partial_{\rho}\phi_1+ \frac{1}{2} \left(\eth \phi_0 - \Bar{\eth} \phi_2 \right) =0. \label{Constraint-Maxwell}
\end{equation}

\begin{remark}
\label{Remark:BoostedData}
{\em Now, consistent with equation \eqref{MaxwellFieldExpansion} in Assumption \ref{Maxwell-field-expansion}  one has that 
\begin{equation}
    \phi_n|_{\mathcal{S}_\star} = \sum_{l=|n-1|}^{\infty} \sum_{m=-l}^l a_{n;l,m}(0) {}_{n-1}Y_{l,m} + o(\rho).
    \label{Maxwell-initial-data}
\end{equation}
Initial data of this type will be called \emph{boosted}. Boosted data can be obtained from \emph{non-boosted} data of the form
\begin{eqnarray*}
&& \phi_1|_{\mathcal{S}_\star} =0, \nonumber\\ 
&& \phi_1|_{\mathcal{S}_\star} = a_{1;0,0,0}(0) Y_{0,0} + o(\rho), \nonumber\\
&& \phi_2|_{\mathcal{S}_\star} =0,
\end{eqnarray*}
via a boost at infinity of the initial hypersurface $\mathcal{S}_\star$.}
\end{remark}

\begin{remark}
\label{Notation}
{\em In the following, we will omit the ${}_{;l,m}$ labels for the coefficients of the field expansion, e.g. $a_1 \equiv a_{1;l,m}$ etc. For initial data, we write $ (a_1)_* \equiv a_{1;l,m}(0)$ etc. }
\end{remark}

For $l \geq 1$, the expansion given by equation \eqref{Maxwell-initial-data} can be used to write equations \eqref{Evolution-Maxwell2} and \eqref{Constraint-Maxwell} at $\tau=0$ as
\begin{subequations}
\begin{equation}
    (\dot{a}_{1})_* + \frac{1}{2} \sqrt{l (l+1)} \left( (a_{0})_* - (a_{2})_* \right) =0, \label{Maxwell-Evolution}
\end{equation}
\begin{equation}
    \frac{1}{2} \sqrt{l (l+1)} \left( (a_{0})_* + (a_{2})_* \right) =0. \label{Maxwell-Constraint}
\end{equation}
\end{subequations}
Thus, {\em for even $l \geq 1$, the condition $(\dot{a}_1)_* =0$ translates into $(a_0)_*=(a_2)_*=0$. }

\begin{remark}
{\em For $l=0$, the condition $\dot{a}_{1,0;0,0}=0$ gives the well-known fact that the Coulomb charges are conserved.}
\end{remark}

A systematic way to obtain solutions to the constraint equations with an expansion of the form given by equation \eqref{Maxwell-initial-data} is to make use of the theory developed in \cite{Andre14}. One has that a symmetric spinor $\phi_{AB}$ satisfying the constraint \eqref{Maxwell-Constraint} can be written as 
\begin{equation}
    \phi_{AB} = (\mathcal{G}_2 \Phi)_{AB} \equiv 2 D_{(A}{}^C \Phi_{B)C},
    \label{Maxwell-spinor-hertzPotential}
\end{equation}
where $D_{AB}$ denotes the spinorial counterpart of the Levi-Civita connection associated with the intrinsic metric on the initial hypersurface $\mathcal{S}_\star$. Consistent with equation \eqref{Maxwell-initial-data} we consider a spinorial potential $\Phi_{AB}$ with components of the form 
\[
     \Phi_n = \sum_{l=|n-1|}^{\infty} \sum_{m=-l}^l b_{n;l,m}(0) {}_{n-1}Y_{l,m} + o(\rho).
\]
A direct computation then gives the following:
\begin{lemma}
The solution given in Proposition \ref{Jacobi-solution} is regular if and only if
\begin{itemize}
    \item[(i)] $(b_1)_*=0$ for even $l$, and
    \item[(ii)] $(b_0)_*=(b_2)_*$ for odd $l$,
\end{itemize}
where $(b_0)_*\equiv b_{0;l,m}(0)$, $(b_1)_*\equiv b_{1;l,m}(0)$ and $(b_2)_*\equiv b_{2;l,m}(0)$.
\label{Maxwell-conditions}
\end{lemma}

Thus, one can always choose freely specifiable initial data for the Maxwell spinor in such a way that the coefficients $a_{1;l,m}$ in expansion \eqref{MaxwellFieldExpansion} are regular at the critical sets $\mathcal{I}^\pm$. As it will be seen in the next subsection this regularity condition is a necessary requirement for the BMS charges to be well-defined at $\mathcal{I}^\pm$. 


\subsection{The Maxwell charges}
\label{Section:The Maxwell charges}
Let in the following $\tilde{F}_{ab}$ denote the physical Maxwell field tensor. Then the source-free Maxwell equations can be written as 
\[
\tilde{\nabla}_b \tilde{F}^{ba} = 0 \hspace{3pt}, \qquad  \tilde{\nabla}_{[a} \tilde{F}_{bc]}=0.
\]
These equations are invariant under the conformal transformation $\bmg= \Omega^2 \tilde{\bmg}$ if one sets
\[
F_{ab} = \tilde{F}_{ab}.
\]
The spinorial counterpart of the Maxwell Field tensor $F_{ab}$ is related to the Maxwell spinor $\phi_{AB}$ via 
\[
F_{AA'BB'} = \phi_{AB} \bmepsilon_{A'B'} + \Bar{\phi}_{A'B'} \bmepsilon_{AB}.
\]
Now, we define the self-dual Faraday tensor $\mathcal{F}_{ab}$ as
\[
\mathcal{F}_{ab} = F_{ab} + \mathrm{i} (*F)_{ab},
\]
where $(*F)_{ab}$ is the Hodge dual of $F_{ab}$ with respect to the volume form of the unphysical metric $\bmg$.  

\medskip
The asymptotic symmetries at null infinity are associated with smooth functions $\lambda$ on $\mathbb{S}^2$ sphere \cite{Kartik18,BarTro11}. For each of the functions $\lambda$, the corresponding charge $\mathscr{Q}$ defined on some cross-section $\mathcal{C}$ of $\mathscr{I}$ is given by
\begin{equation}
    \mathscr{Q} \equiv \int_C \lambda \mathcal{F}_{ab} l^a n^b \mathrm{d}S,
    \label{Charge-Maxwell}
\end{equation}
where $l^a$ and $n^a$ are elements of the NP-gauge null tetrad and $\mathrm{d}S$ denotes the area element of the cross-section $\mathcal{C}$. We make the identification $l^a \equiv \bme'_{\bmzero
\bmzero'}$ and $n^a \equiv \bme'_{\bmone \bmone'}$ at $\mathscr{I}^+$.Thus, in terms of the Maxwell spinor $\phi_{AB}$, one can verify that
\[
\mathscr{Q} = 2 \int_C \lambda \Bar{\phi}_1 \mathrm{d}S,
\]
with the understanding that $\Bar{\phi}_1$ is evaluated at future null infinity. The electric and magnetic parts of the charge are then given by
\begin{eqnarray*}
&& \mathscr{Q}_{[e]} \equiv - \mathrm{Re}\,[\mathscr{Q}], \\
&& \mathscr{Q}_{[m]} \equiv \mathrm{Im}\,[\mathscr{Q}].
\end{eqnarray*}
Spherical harmonics $Y_{l,m} \equiv {}_{0}Y_{l,m}$ form a complete basis on $\mathbb{S}^2$. Thus, for simplicity, one can choose $\lambda = Y_{l,m}$. Then, using the expansion given by Assumption \ref{Maxwell-field-expansion}, one can show the BMS charges at $\mathcal{I}^+$ ($\tau=1$, $\rho=0$) can be written as
\[
\mathscr{Q}|_{\mathcal{I}^+} = 2 \Bar{a}_{1;l,m}(1).
\]
In the case of past null infinity $\mathscr{I}^-$, the roles of $l^a$ and $n^a$ in \eqref{Charge-Maxwell} are reversed. In view of  the symmetry of equation \eqref{Charge-Maxwell} in $l^a$ and $n^a$ one readily finds that at $\mathcal{I}^-$ ($\tau=-1$, $\rho=0$) one has
\[
\mathscr{Q}|_{\mathcal{I}^-} = 2 \Bar{a}_{1;l,m}(-1).
\]
Then, for generic Cauchy initial data for the Maxwell equations, one can deduce that the BMS charges at $\mathcal{I}^{+}$ are not well-defined unless the free specifiable data satisfy the conditions given
in Lemma \ref{Maxwell-conditions} so that the coefficients $a_{1;l,m}(\tau)$ contain no logarithmic singularity ---see Remark \ref{Remark:Logs}. A similar argument applies for the charges at $\mathcal{I}^-$. Making use of the results in Section \ref{Section:Initial-data-for-Maxwell}, one can show: 

\begin{theorem}
\label{Theorem:ChargesMaxwell}
Given initial conditions for the Maxwell equations satisfying the conditions in Lemma \ref{Maxwell-conditions} so that $B_{l,m}$ vanishes, the asymptotic charges for the Maxwell field on Minkowski spacetime can be expressed in terms of the initial data on some initial hypersurface. One has that:
\begin{itemize}
\item[(i)] at $\mathcal{I}^+$, the charges are
given by
\begin{equation}
    \mathscr{Q}|_{\mathcal{I}^+} = \begin{cases} 
             -2 (l+1) Q_{l+1}(0) (a_1)_* \qquad \text{for even }l \geq 0, \\
             \sqrt{l(l+1)} Q_{l}(0) \left( (a_0)_* - (a_2)_* \right) \qquad \text{for odd }l.
        \end{cases}
    \label{TheMaxwellCharge}
\end{equation}
\item[(ii)] At $\mathcal{I}^-$, one obtains
\begin{equation}
    \mathscr{Q}|_{\mathcal{I}^-} = \begin{cases} 
             - 2 (l+1) Q_{l+1}(0) (a_1)_* \qquad \text{for even }l \geq 0, \\
             - \sqrt{l(l+1)} Q_{l}(0) \left( (a_0)_* - (a_2)_* \right) \qquad \text{for odd }l.
        \end{cases}
    \label{TheMaxwellCharge2}
\end{equation}
\end{itemize}
If the conditions in Lemma 2 are not satisfied, then the BMS charges are not well-defined at $\mathcal{I}^\pm$. 
\end{theorem}

\begin{remark}
{\em The charges can be expressed in terms of freely specifiable data using equation  \eqref{Maxwell-spinor-hertzPotential}. For example, at $\mathcal{I}^+$ we get
 \[
    \mathscr{Q}|_{\mathcal{I}^+} = 2\sqrt{2} l(l+1) Q_l(0) (b_1)_* \hspace{6mm} \text{for odd } l.
 \]
  One can obtain expressions for $\mathscr{Q}|_{\mathcal{I}^+}$ for even $l$ in a similar fashion.}
\end{remark}

\begin{remark}
{\em Notice that the charges at $\mathcal{I}^-$ are equal to charges at $\mathcal{I}^-$ multiplied by $(-1)^l$ factor. More specifically, the charges at $\mathcal{I}^{\pm}$ are written in terms of the same piece of initial data so that the vanishing of the initial data for a given $l$ implies the vanishing of both charges.} 
\end{remark}


\section{The spin-2 field and the asymptotic charges in the F-gauge}
\label{Section:Spin2}

In this section, we discuss spin-2 fields on Minkowski spacetime and carry out an analysis analogous to that of the Maxwell field in the previous section. We start by writing the spin-2 field in terms of spin-weighted spherical harmonics and expanding the field equations in terms of the F-gauge. Finally, we obtain a solution for the coefficient of the expansion that appears in the definition of the asymptotic charges. 
\medskip

\subsection{The spin-2 equation}
Let $\psi_{ABCD}=\psi_{(ABCD)}$ denote a valence 4 symmetric spinor. The massless spin-2 field equation in spinorial formalism can be written as
\begin{equation}
    \nabla^A{}_{A'} \psi_{ABCD} = 0.
    \label{Spin2-field-equation}
\end{equation}
Applying $-2 \nabla_{C}{}^{A'}$, one readily obtains the wave equation for the spin-2 field:
\begin{eqnarray}
\square \psi_{ABCD} = 0. \label{spin-2}
\end{eqnarray}
Using the F-gauge spin-dyad, we define the components of the spin-2 field as
\begin{eqnarray*}
& \psi_0 \equiv \psi_{ABCD} o^A o^B o^C o^D, \nonumber\\
& \psi_1 \equiv \psi_{ABCD} o^A o^B o^C \iota^D,& \nonumber\\
& \psi_2 \equiv \psi_{ABCD} o^A o^B \iota^C \iota^D,\nonumber\\
& \psi_3 \equiv \psi_{ABCD} o^A \iota^B \iota^C \iota^D, \nonumber\\
& \psi_4 \equiv \psi_{ABCD} \iota^A \iota^B \iota^C \iota^D. 
\end{eqnarray*}
Expanding equation \eqref{spin-2}, it follows that in the F-gauge, one has
\begin{subequations}
\begin{eqnarray}
&& (1-\tau^2) \ddot{\psi}_0 + 2 \rho \tau \partial_{\rho}\dot{\psi}_0- \rho^2 \partial^2_{\rho}\psi_0 - \frac{1}{2} \left( \eth \Bar{\eth} \psi_0 + \Bar{\eth} \eth \psi_0 \right) + 2 (2-\tau) \dot{\psi}_0 + 4 \psi_0 =0, \label{WaveEqnSpin2Expanded-1} \\ 
&& (1-\tau^2) \ddot{\psi}_1 + 2 \rho \tau \partial_{\rho}\dot{\psi}_1- \rho^2 \partial^2_{\rho}\psi_1 - \frac{1}{2} \left( \eth \Bar{\eth} \psi_1 + \Bar{\eth} \eth \psi_1 \right) + 2 (1-\tau) \dot{\psi}_1 + \psi_1 =0, \label{WaveEqnSpin2Expanded-2} \\ 
&& (1-\tau^2) \ddot{\psi}_2 + 2 \rho \tau \partial_{\rho}\dot{\psi}_2- \rho^2 \partial^2_{\rho}\psi_2 - \frac{1}{2} \left( \eth \Bar{\eth} \psi_2 + \Bar{\eth} \eth \psi_2 \right) - 2 \tau \dot{\psi}_2 =0, \label{WaveEqnSpin2Expanded-3} \\ 
&& (1-\tau^2) \ddot{\psi}_3 + 2 \rho \tau \partial_{\rho}\dot{\psi}_3- \rho^2 \partial^2_{\rho}\psi_3 - \frac{1}{2} \left( \eth \Bar{\eth} \psi_3 + \Bar{\eth} \eth \psi_3 \right) - 2 (1+\tau) \dot{\psi}_3 + \psi_3=0, \label{WaveEqnSpin2Expanded-4} \\ 
&& (1-\tau^2) \ddot{\psi}_4 + 2 \rho \tau \partial_{\rho}\dot{\psi}_4- \rho^2 \partial^2_{\rho}\psi_4 - \frac{1}{2} \left( \eth \Bar{\eth} \psi_4 + \Bar{\eth} \eth \psi_4 \right) - 2 (2+\tau) \dot{\psi}_4 + 4 \psi_4=0. \label{WaveEqnSpin2Expanded-5}
\end{eqnarray}
\end{subequations}

To analyse the solutions for these equations, we make the following Ansatz:

\begin{assumption}
\label{spin-2-field-expansion}
The components of the spin-2 field can be expanded near $\rho=0$ and $\tau\in[-1,1]$ as
\begin{eqnarray}
\psi_n = \sum_{l=|2-n|}^\infty \sum_{m=-l}^l a_{n;l,m}(\tau) {}_{2-n}Y_{l,m} + o_1(\rho), \label{expansion-spin-2-field}
\end{eqnarray}
where $a_{n;l,m}: \mathbb{R} \rightarrow \mathbb{C}$ and $n=0,...,4$. 
\end{assumption}

\begin{remark}
{\em As in the case of the Maxwell field, making use of the estimates developed in \cite{Fri03b}, it is possible to show that there exist large classes of solutions to the spin-2 equations having an expansion consistent with
  \eqref{expansion-spin-2-field}. More precisely, one has
  \[
  \psi_n = \sum_{k=|1-k|}^{p-1} \frac{1}{k!}\psi_n^{(k)}\rho^k + R_p[\psi_n],
  \]
  where the coefficients $\psi_n^{(k)}$ are explicitly known functions of $\tau$ and the angular variables which are smooth for $\tau\in(-1,1)$ and whose regularity at $\tau=\pm 1$ can be controlled in terms of the initial data. Moreover, the reminder satisfies $R_p[\psi_n]\in C^m$ for $p\geq m+6$ for $\rho$ near 0 and $\tau\in [-1,1]$. In particular, for expansion \eqref{expansion-spin-2-field} in Assumption \ref{spin-2-field-expansion} one requires $p\geq 7$.}
\end{remark}

Using the expansion Ansatz \eqref{expansion-spin-2-field} and
\eqref{WaveEqnSpin2Expanded-1}-\eqref{WaveEqnSpin2Expanded-5}, one gets
\begin{subequations}
\begin{equation}
    (1-\tau^2) \ddot{a}_{0;l,m} + 2 (2-\tau) \dot{a}_{0;l,m} + l (l+1) a_{0;l,m} =0, \label{WaveEqnSpin2Expansion1} \\
\end{equation}
\begin{equation}
    (1-\tau^2) \ddot{a}_{1;l,m} + 2(1-\tau) \dot{a}_{1;l,m} +l (l+1) a_{1;l,m} =0, \label{WaveEqnSpin2Expansion2} \\
\end{equation}
\begin{equation}
    (1-\tau^2) \ddot{a}_{2;l,m} - 2\tau \dot{a}_{2;l,m} + l (l+1) a_{2;l,m} =0, \label{WaveEqnSpin2Expansion3}  \\
\end{equation}
\begin{equation}
    (1-\tau^2) \ddot{a}_{3;l,m} -2(1+\tau) \dot{a}_{3;l,m} + l (l+1) a_{3;l,m}=0, \label{WaveEqnSpin2Expansion4} \\
\end{equation}
\begin{equation}
    (1-\tau^2) \ddot{a}_{4;l,m} - 2(2+\tau) \dot{a}_{4;l,m} + l (l+1) a_{4;l,m}=0. \label{WaveEqnSpin2Expansion5}
\end{equation}
\end{subequations}
We are interested in obtaining a solution to equation \eqref{WaveEqnSpin2Expansion3} as the asymptotic BMS charges for the spin-2 field are expressed in terms of $a_{2;l,m}$. The solution to equation \eqref{WaveEqnSpin2Expansion3} is similar to the solution to \eqref{WaveEqnExpansion2} in the Maxwell case. Thus, we have the following:

\begin{proposition}
For $ l \geq 0$ and $-l \leq m \leq l$, the solution to \eqref{WaveEqnSpin2Expansion3} is given by
$$
a_{2;l,m} = A_{l,m} P_{l}(\tau) + B_{l,m} Q_{l}(\tau) 
$$
where $A_{l,m}$ and $B_{l,m}$ are constants that can be expressed in
terms of initial conditions $a_{2;l,m}(0)$ and $\dot{a}_{2;l,m}(0)$ as
\begin{subequations}
\begin{equation}
    A_{l,m} = -(l+1) Q_{l+1}(0) a_{2;l,m}(0) - Q_{l}(0) \dot{a}_{2;l,m}(0), \label{Spin2-Alm}
\end{equation}
\begin{equation}
        B_{l,m} = \frac{\sqrt{\pi} (l+1)}{\Gamma(-\frac{l}{2}) \Gamma(\frac{l+3}{2})} a_{2;l,m}(0) + \frac{\sqrt{\pi}}{\Gamma(\frac{1}{2}-\frac{l}{2}) \Gamma(1+\frac{l}{2})} \dot{a}_{2;l,m}(0). \label{Spin2-Blm}
\end{equation}
\end{subequations}
\label{Solution-spin-2}
\end{proposition}

Similar to the case of Maxwell equations, the solution given by Proposition \ref{Solution-spin-2} diverges near $\tau = \pm 1$ if $B_{l,m}\neq 0$. Thus, we require $B_{l,m}$ to be vanishing to get a regular solution at the critical sets. Let $(a_2)_* \equiv a_{2;l,m}(0)$ and $(\dot{a}_2)_* \equiv \dot{a}_{2;l,m}(0)$, one can show the following:

\begin{lemma}
\label{Initial-data-Spin2}
The solution in Proposition \ref{Solution-spin-2} is regular if and only if:
\begin{itemize}
\item[(i)] $(a_2)_*=0$ for odd $l$, and
\item[(ii)] $(\dot{a}_2)_*=0$ for even $l$.
\end{itemize}
\end{lemma}


\subsection{Initial data for the spin-2 equations}
\label{Section:Initial-data-spin2}
The spin-2 field equation implies a relation between the coefficients of the expansions given in equation \eqref{expansion-spin-2-field}. In this section, our aim is to derive the conditions on the freely
specifiable data so that the solution for equation \eqref{WaveEqnSpin2Expansion3} is non-diverging. One can decompose equation \eqref{Spin2-field-equation} to obtain the evolution and constraint equations as follows:
\begin{subequations}
\begin{equation}
    \mathcal{D} \psi_{ABCD} + 2 \mathcal{D}_A{}^E \psi_{EBCD} =0, \hspace{3mm} \text{(evolution equations)}
\end{equation}
\begin{equation}
    \mathcal{D}^{AB}\psi_{ABCD} =0. \hspace{21.5mm} \text{(constraint
   equation)} \label{GaussConstraint}
\end{equation}
\end{subequations}
Expanding in terms of the F-gauge, the evolution equations for the spin-2 field are given by
\begin{subequations}
\begin{equation}
    (1+\tau) \dot{\psi}_0 - \rho \partial_{\rho}\psi_0 - \eth \psi_1 = -2 \psi_0, \label{EvolutionEqnSpin2-1} \\
\end{equation}
\begin{equation}
    (1+\frac{\tau}{2}) \dot{\psi}_1 - \frac{\rho}{2} \partial_{\rho}\psi_1 - \frac{\Bar{\eth} \psi_0}{4} - \frac{3 \eth \psi_2}{4} = - \psi_1, \label{EvolutionEqnSpin2-2}\\
\end{equation}
\begin{equation}
    \dot{\psi}_2 - \frac{\Bar{\eth} \psi_1}{2} - \frac{\eth \psi_3}{2} =0, \label{EvolutionEqnSpin2-3}\\
\end{equation}
\begin{equation}
    (1-\frac{\tau}{2}) \dot{\psi}_3 + \frac{\rho}{2} \partial_{\rho}\psi_3 - \frac{3 \Bar{\eth} \psi_2}{4} - \frac{\eth \psi_4}{4} = \psi_3, \label{EvolutionEqnSpin2-4} \\
\end{equation}
\begin{equation}
    (1-\tau) \dot{\psi}_4 + \rho \partial_{\rho}\psi_4- \Bar{\eth}\psi_3 = 2 \psi_4, \label{EvolutionEqnSpin2-5}
\end{equation}
\end{subequations}
and the constraint equations are given by
\begin{subequations}
\begin{equation}
    \tau \dot{\psi}_1 - \rho \partial_{\rho}\psi_1 + \frac{\Bar{\eth} \psi_0}{2} - \frac{\eth \psi_2}{2} =0, \label{ConstraintEqnSpin2-1}  \\
\end{equation}
\begin{equation}
    \tau \dot{\psi}_2 - \rho \partial_{\rho}\psi_2 + \frac{\Bar{\eth}\psi_1}{2} - \frac{\eth \psi_3}{2} =0, \label{ConstraintEqnSpin2-2}  \\
\end{equation}
\begin{equation}
     \tau \dot{\psi}_3 - \rho \partial_{\rho}\psi_3 + \frac{\Bar{\eth}\psi_2}{2} - \frac{\eth \psi_4}{2}=0. \label{ConstraintEqnSpin2-3} 
\end{equation}
\end{subequations}

Now, we consider an expansion which is consistent with Assumption \ref{spin-2-field-expansion} so that $\psi_n|_{\mathcal{S}_\star}$ takes the form
\begin{equation}
    \psi_n|_{\mathcal{S}_\star} = \sum_{l=|2-n|}^{\infty} \sum_{m=-l}^{l} a_{n;l,m}(0) {}_{2-n} Y_{l,m} + o(\rho).
    \label{spin2-initial-data}
\end{equation}
As in the case of the Maxwell field, we call this class of initial data \emph{boosted} ---see Remark \ref{Remark:BoostedData}. Making use of this expansion, equations \eqref{EvolutionEqnSpin2-3} and
\eqref{ConstraintEqnSpin2-2} evaluated at $\tau=0$, yield
\begin{eqnarray*}
&& (\dot{a}_{2})_* + \frac{1}{2} \sqrt{l (l+1)} \left( (a_{3})_* - (a_{1})_* \right) =0, \nonumber \\
&& \frac{1}{2} \sqrt{l(l+1)} \left( (a_{3})_* + (a_{1})_* \right) =0,
\end{eqnarray*}
where the index ${}_{;l,m}$ have been omitted and we use the notation introduced in Remark \ref{Notation}. In particular, from the above one can deduce that{ \em at $\tau=0$, the condition $(\dot{a}_2)_*=0$ translates to $(a_1)_*=(a_3)_*=0$ for even $l \geq 2$.}

\medskip
In order to write the initial conditions in terms of freely specifiable data, we make use of the theory developed in \cite{Andre14} for solutions to the Gauss constraint
\eqref{GaussConstraint}. Specifically, for a spin-2 field satisfying the constraint equation, one can write
\begin{equation}
    \psi_{ABCD} = (\mathcal{G}_4 \Psi)_{ABCD} = 2 \left( \stwist_2{\sdiv_4{\scurl_4{\Psi}}} \right)_{ABCD} + 8 \left( \scurl_4{\scurl_4{\scurl_4{\Psi}}} \right)_{ABCD},
    \label{Hertz-potential-spin2} 
\end{equation}
where $\sdiv_4, \scurl_4$ and $\stwist_2$ are the divergence, curl and twistor operators as defined in \cite{Andre14}. Consistent with the previous discussion, we assume that the components of the Hertz potential $\Psi_{ABCD}$ can be expanded in terms of spin-weighted spherical harmonics as
\begin{equation}
    \Psi_n = \sum_{l=|2-n|}^\infty \sum_{m=-l}^l  b_{n;l,m}(0) {}_{2-n }Y_{l,m} + o(\rho).
    \label{Hertz-Potential-Spin2-expansion}
\end{equation}

Letting $(b_0)_*\equiv b_{0;l,m}(0)$, $(b_1)_*\equiv b_{1;l,m}(0)$, $(b_2)_*\equiv b_{2;l,m}(0)$, $(b_3)_*\equiv b_{3;l,m}(0)$ and $(b_4)_*\equiv b_{4;l,m}(0)$, one can expand equation
\eqref{Hertz-potential-spin2} in  the F-gauge and use \eqref{expansion-spin-2-field} and \eqref{Hertz-Potential-Spin2-expansion} to obtain:
\begin{lemma}
The solution given in Proposition \ref{Solution-spin-2} is regular if and only if:
\begin{itemize}
    \item[(i)] $(b_2)_* = C_l (b_4)_* - D_l (b_0)_*$ for even $l \geq 1$, and
    \item[(ii)] $(b_1)_*=(b_3)_*$ for odd $l$,
\end{itemize}
where $C_l$ and $D_l$ are constants that depend on $l$ and whose specific form is not relevant for the subsequent discussion.
\label{Spin2-conditions}
\end{lemma}

\begin{remark}
{\em It follows then that, as in the case of the Maxwell spin field, the conditions ensuring solutions that are regular at the critical sets can be encoded in terms of freely specifiable data.}
\end{remark}


\subsection{The asymptotic BMS charges for the spin-2 field}
In this section, we consider the asymptotic charges for the spin-2 field associated with supertranslations. For the spin-2 field, the asymptotic symmetries at null infinity are given by the smooth functions $\lambda$ on $\mathbb{S}^2$. The associated BMS charges defined on some cross-section $\mathcal{C}$ of null infinity can be written as
\[
\mathscr{P} = \int_{\mathcal{C}} \lambda \mathcal{W}_{abcd} l^a n^b
m^c \Bar{m}^d \mathrm{d}S,
\]
where $\mathcal{W}_{abcd}$ is given by
$$
\mathcal{W}_{abcd} = W_{abcd} + \mathrm{i} (*W)_{abcd}.
$$
Here, $W_{abcd}$ denotes a Weyl-like tensor. The spinorial counterpart $W_{AA'BB'CC'DD'}$ satisfies
$$
W_{AA'BB'CC'DD'} = -\psi_{ABCD} \bmepsilon_{A'B'} \bmepsilon_{C'D'} - \Bar{\psi}_{A'B'C'D'} \bmepsilon_{AB} \bmepsilon_{CD}.
$$
The BMS charges can then be written in terms of the spin-2 spinor $\psi_{ABCD}$. At $\mathscr{I}^+$, if one makes the identification $l^a \equiv \bme'_{\bmzero \bmzero'}$, $n^a \equiv \bme'_{\bmone
\bmone'}$ and $m^a \equiv \bme_{\bmzero \bmone'}$, then the charges can be written as
$$
\mathscr{P} = - 2 \int_{\mathcal{C}} \lambda \Bar{\psi}_2 \mathrm{d}S.
$$
Using the expansion given in Assumption \ref{spin-2-field-expansion} and assuming $\lambda = Y_{l,m}$, the expression for the charges at $\mathcal{I}^+$ can be written as
$$
\mathscr{P}|_{\mathcal{I}^+} = -2 \Bar{a}_{2;l,m}(1).
$$
At $\mathcal{I}^{-}$, one can show that 
$$
\mathscr{P}|_{\mathcal{I}^-} = -2 \Bar{a}_{2;l,m}(-1).
$$
Making use of the results of Section \ref{Section:Initial-data-spin2}, one readily obtains the following:

\begin{theorem}
\label{Theorem:ChargesSpin2}
Given initial conditions for the spin-2 field equations satisfying Lemma \ref{Spin2-conditions} so that $B_{l,m}$ vanishes, the asymptotic charges for the spin-2 field on Minkowski spacetime can be expressed in terms of the initial data on some initial hypersurface.  More precisely, one has that:
\begin{itemize}
\item[(i)] At $\mathcal{I}^+$, the charges are given by
\[
\mathscr{P}|_{\mathcal{I}^+} = \begin{cases} 
         2 (l+1) Q_{l+1}(0) (a_2)_* \qquad \text{for even }l \geq 0, \\
         \sqrt{l(l+1)} Q_{l}(0) \left( (a_1)_* - (a_3)_* \right) \qquad \text{for odd }l.
    \end{cases}
\]
\item[(ii)] At $\mathcal{I}^-$, the charges are given by
\[
\mathscr{P}|_{\mathcal{I}^-} = \begin{cases} 
         2 (l+1) Q_{l+1}(0) (a_2)_* \qquad \text{for even }l \geq 0, \\
         - \sqrt{l(l+1)} Q_{l}(0) \left( (a_1)_* - (a_3)_* \right) \qquad \text{for odd }l.
    \end{cases}
\]
\end{itemize}
If the conditions of Lemma \ref{Spin2-conditions} are not satisfied then the BMS charges are not well-defined.
\end{theorem}

\begin{remark}
{\em As in the case of the Maxwell field, the BMS asymptotic charges for the spin-2 field can be fully expressed in terms of freely specifiable data using equation  \eqref{Hertz-potential-spin2}. The resulting expressions are much more involved and not particularly illuminating.}
\end{remark}

\begin{remark}
{ \em Similar to the case of the Maxwell field, for a given harmonic $Y_{l,m}$, the charges at $\mathcal{I}^+$ and $\mathcal{I}^-$ can be traced back to the same piece of freely specifiable data. In this sense, one can say they contain the same information. }
\end{remark}


\section{Conclusions}
In this article, we have analysed the relation between the asymptotic charges for the Maxwell and spin-2 fields at $\mathcal{I}^{\pm}$ and initial data given on some Cauchy hypersurface $\mathcal{S}_\star$ of the Minkowski spacetime. The study of the asymptotic charges in GR is of physical importance as it is directly related to the continuous symmetries of the system under consideration. In our analysis, we have focused on charges associated with smooth functions defined on $\mathbb{S}^2$. In full General Relativity, these charges would correspond to the supertranslation charges of the BMS group. 

\medskip
For the Maxwell field, our analysis shows that the asymptotic charges are not in general defined unless one imposes extra conditions on the initial data. To show this, we expanded the initial data in terms of spin-weighted spherical harmonics, and we showed that the asymptotic charges at $\mathcal{I}^{+}$ are well-defined and are fully written in terms of the initial data if and only if the freely specifiable initial data satisfy the conditions given in Lemma \ref{Maxwell-conditions}. Similarly, the asymptotic charges at $\mathcal{I}^-$ can be written in terms of the freely specifiable initial data on $\mathcal{S}$ and we show that they correspond to the same initial data used for the asymptotic charges at $\mathcal{I}^+$.

\medskip
A similar analysis with analogous conclusions was carried out for the spin-2 field. In particular, we show that the asymptotic charges at $\mathcal{I}^{\pm}$ are well-defined and can be written in terms of
the freely specifiable initial data if and only if the conditions given in Lemma \ref{Spin2-conditions} are satisfied. Given these conditions, we show that there is a correspondence between the charges
at $\mathcal{I}^+$ and $\mathcal{I}^-$ in the sense that both charges are written in terms of the same initial data.

\medskip
In summary, our analysis shows that the asymptotic BMS charges at $\mathcal{I}^{\pm}$ are not defined for generic initial data unless we choose initial data that satisfy the conditions given in Lemma \ref{Maxwell-conditions} in the Maxwell case and Lemma \ref{Spin2-conditions} in the spin-2 case. Moreover, the resulting expressions for the charges in terms of the freely specifiable data show that, for a given harmonic, there is a natural correspondence between the future and past charges without the need for any further assumptions. The analysis in this article relies on the leading order structure of the cylinder at spatial infinity ---i.e. its Minkowskian part. As this is a universal aspect of the structure of spatial infinity for more general spacetimes (e.g. the Schwarzschild spacetime or, more generally, any stationary spacetime) it is conjectured that the results in this article can be extended to massless test fields propagating on stationary spacetimes ---the particular details, however, need to be verified. For more general (i.e. non-stationary) spacetimes, the situation is less clear because of the generic non-smooth behaviour of the spacetime near spatial infinity.

The approach followed in this article can be generalised to the case of the asymptotic BMS charges on asymptotically flat spacetimes. The general properties of the asymptotic expansions of the conformal Einstein field equations required for this type of analysis have been studied in \cite{Fri98a,FriKan00}. Notice, however, that the analysis in these two references is restricted to the case of spacetimes arising from time-symmetric initial data conditions ---i.e. vanishing extrinsic curvature. Thus, an important part of the extension of the analysis in the present article to the vacuum Einstein field equations (or the Einstein-Maxwell system) is to obtain a suitably large class of initial data sets for the Einstein field equations giving rise to non-vanishing BMS charges. The analysis for the Einstein vacuum field equations will be discussed elsewhere.


\section{Acknowledgements}
M.M.A.Mohamed would like to thank Mahdi Godazgar for the useful discussions while working on this project. We also thank Kartik Prabhu for interesting conversations. Some calculations in this project used the computer algebra system Mathematica with the package xAct for tensor manipulations ---see \cite{xAct}.


\end{document}